\documentclass[prl,twocolumn,epsf,psfig]{revtex4}
\usepackage{graphicx}
\DeclareGraphicsExtensions{.pdf,.png,.gif,.jpg}

\begin{document}

\title{Compressibility and entropy of cold fermions in one dimensional optical lattices}
\author{Andrew Snyder, Iori Tanabe, and Theja De Silva}
\affiliation{Department of Physics, Applied Physics and Astronomy,
The State University of New York at Binghamton, Binghamton, New York
13902, USA.}

\begin{abstract}
We calculate several thermodynamic quantities for repulsively interacting one dimensional fermions. We solve the Hubbard model at both zero and finite temperature using the Bethe ansatz method. For arbitrary values of the chemical potential, we calculate the particle number density, the double occupancy, various compressibilities, and the entropy as a function of temperature and interaction. We find that these thermodynamic quantities show a characteristic behavior so that measurements of these quantities can be used as a detection of temperature, the metal-insulator transition, and metallic and insulating phases in the trap environment. Further, we discuss an experimental scheme to extract these thermodynamic quantities from the column density profiles. The entropy and the compressibility of the entire trapped atomic cloud also reveal characteristic features indicating whether insulating and/or metallic phases coexist in the trap.

\end{abstract}
\maketitle
\section{I. Introduction}

Based on Bethe ansatz solutions of the spin-half Heisenberg chain~\cite{bethe}, Lieb and Wu proposed solutions of the one dimensional(1D) Hubbard model as a set of algebraic equations, known as the Lieb-Wu equations~\cite{lieb}. Since then, the 1D Hubbard model has been considered as an exactly solvable model and is used as a toy model for non-perturbative strongly correlated effects in interacting electron systems. Later, Takahashi developed a classification of the Lieb-Wu equations and introduced a set of non-linear integral equations, known as the thermodynamic Bethe ansatz (TBA), to replace the Lieb-Wu equations. Any thermodynamic quantity that relates to the energy spectrum of the 1D Hubbard model can be calculated from numerical solutions of the TBA equations~\cite{revHM}. Due to the advent of Feshbach resonances and optical lattices in the field of ultra-cold atoms~\cite{revCG}, the 1D Hubbard model has evolved from a toy model to a paradigm of experimental relevance for strongly correlated particle systems. The high degree of tunability and the absence of disorder in optical lattice experiments offer an opportunity to explore fundamental physical phenomena in a controlled manner.

The first successful experimental emulation of the Hubbard model in optical lattices was achieved by Greiner et al.~\cite{greiner} using bosonic atoms. This was after the theoretical prediction of a Bose-superfluid to Mott-insulator quantum phase transition by Jaksch \emph{et al}~\cite{jaksch}. With recent extraordinary experimental developments in manipulating Fermi atoms in optical lattices~\cite{fermi1, fermi2, fermi3, fermi4}, experimentalists are now addressing the rich many-body scenario in these systems. Fermions in optical lattices can open new regimes and routes for understanding many body physics beyond condensed matter models. For example, mixtures of two-component atoms with different masses introduce spin-dependent hopping amplitudes. This may effect the stability of the possible quantum phases or even induce new phases.

Parallel to the experimental progress of manipulating fermions in optical lattices, recently there have been theoretical studies on the role of an external harmonic potential and its effect on compressibility and double occupancies~\cite{t1, t2, t3, t4, t5, t6}. The underlying harmonic trapping potential present in all cold-gas experiments causes density to vary across the lattice. This trapping-potential induced inhomogeneity allows for the spatial coexistence of metallic and insulating phases. Measurements of local and/or global thermodynamic properties allows one to detect these different phases. In the present study, we use an exactly solvable 1D Hubbard model with the local density approximation to extract the thermodynamic quantities in an inhomogeneous system. Our results will contribute to the worldwide experimental efforts in simulating theoretical models of interacting systems using cold atoms in optical lattices. In this paper, we calculate local thermodynamic quantities such as local density, number of double occupation sites, entropy, and various compressibilities for the 1D Fermi-Hubbard model. We present our results as a function of the chemical potential, particle number density and total number of atoms. The chemical potential and particle number density monotonically vary across the lattice due to the external harmonic potential so that the chemical potential can be considered as an effective spatial coordinate. At zero temperature and high temperatures, our work is complementary to some of the work presented in refs.~\cite{t1, t2, t3, t4, t5, t6, barbi}. However, we go beyond these studies by systematically extending zero temperature calculations to any finite temperature. We use both the Bethe ansatz and thermodynamic Bethe ansatz approaches. We also present an experimental scheme to deduce the calculated thermodynamic quantities from the column densities. Further, using the local density approximation, we calculate the entropy, compressibility, and total number of double occupation sites as a function of the total number of atoms in the lattice.

The present paper is organized as follows. In section II, we present our model. In section III, we present our zero temperature calculation procedure and the results. In section IV, we discuss our finite temperature TBA calculation and present various thermodynamic quantities for representative values of system parameters. In section V, we provide an experimental scheme to show the connection between calculated thermodynamic quantities and experimentally measurable column densities followed by the calculation of thermodynamic quantities in a trapped system. Finally in section VI, we provide a discussion and draw our conclusions.

\section{II. The model}

We consider a two-component, cold Fermi gas in a 1D optical lattice with $N$ atoms and $N_s$ lattice sites. We assume that the inter-site tunneling amplitude $t$ is independent of the pseudospin ($\sigma$). Here the pseudospin $\sigma = \uparrow, \downarrow$ refers to the two hyperfine states of the atoms. The system of fermions in the lattice can be described by the Fermi-Hubbard model:

\begin{eqnarray}
H = - t\sum_{<ij>,\sigma} c^\dagger_{i\sigma}c_{j\sigma}+U\sum_in_{i\uparrow}n_{i\downarrow}
- \mu \sum_{i\sigma} c^\dagger_{i\sigma}c_{i\sigma}
\end{eqnarray}

\noindent where $c^\dagger_{i\sigma} (c_{i\sigma})$ creates (destroys) a Fermi atom with pseudospin $\sigma$ at lattice site $i$. The density operator is $n_{i \sigma}=c^\dagger_{i\sigma} c_{i\sigma}$, $<ij>$ indicates the nearest
neighbor pair of sites and the chemical potential for fermions is $\mu$. In general, the external trapping potential $V_{i} =
\gamma \mathbf{R}_i^2$ at site $i$ for atoms can be treated in the local density approximation by replacing $\mu$ by $\mu-V_{i}$. The on-site interaction $U$ is sensitive to both laser intensity and the \emph{s}-wave scattering length of the two Fermi species. As $U$ is linearly proportional to the \emph{s}-wave scattering length, it can be tuned to negative or positive values by exploiting Feshbach resonances. In the present study, we consider the repulsive Hubbard model so that $U$ can have only positive values. In cold-gas experiments, the tunneling amplitude is controlled by the intensity of the standing laser waves. In current experimental setups, the tunneling amplitude is exponentially sensitive to the laser intensity while the on-site interaction is weakly sensitive. Thus, $U/t$ can be controlled by a single parameter: the laser intensity. Further, standing laser waves allow for a variation of the dimensionality of the system. A reduced dimensionality is achieved by freezing the atomic motion in certain directions. Experimentally, the 1D geometry can be realized by a strong confinement in the transverse direction with an additional periodic potential applied along the other direction. The number of atoms in each pseudospin state is conserved because of the large energy gap between different spin states due to the external magnetic field present in most experiments. In the present study, we consider an equal number of atoms in each pseudospin state.

\section{III. Zero temperature limit}

The 1D ground-state energy of the Fermi-Hubbard model can be calculated from
a Bethe ansatz~\cite{revHM, lw, tak1}. As a very good approximation, the ground-state energy per site $E(n, U, t) = t \times e(n,U/t=u)$ can be written based on the data obtained from a numerical solution of the Bethe-Ansatz~\cite{t1, t2, mfs},

\begin{figure}
\includegraphics[width=\columnwidth]{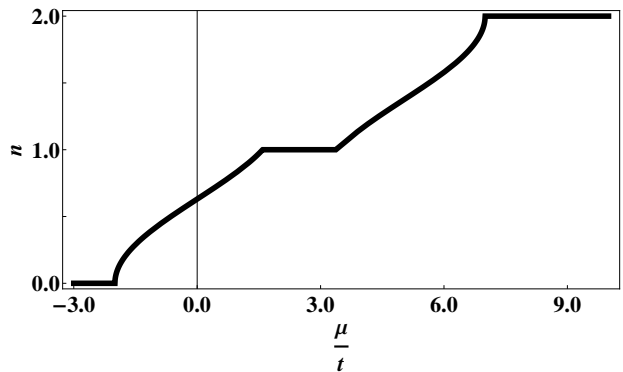}
\caption{Zero-temperature particle density as a function of the chemical potential $\mu$ at $U/t = 5$.} \label{f1}
\end{figure}

\begin{eqnarray}
e_{<}(n, u) = -\frac{2 f(u)}{\pi}\sin\biggr(\frac{\pi n}{f(u)}\biggr).
\end{eqnarray}

\noindent This is valid in the region where $0 \leq n \leq 1$, where $n = N/N_s$ is the average atom density per site. For the region $1 \leq n \leq 2$, the energy can be obtained from the particle-hole transformation; $e_{>}(n, u) = (n-1)u + e_{<}(2-n, u)$. The function $f(u)$ is determined from the equation

\begin{eqnarray}
I \equiv -\frac{f}{\pi}\sin\biggr(\frac{\pi}{f}\biggr) = -2\int_0^\infty\frac{J_0(x)J_1(x)}{x(1 + e^{ux/2})}dx,
\end{eqnarray}

\noindent where $J_m$ is the $m$th order Bessel function. We find the chemical potential by minimizing the grand canonical energy under the fixed total particle numbers; $\mu/t = \partial e/\partial n$,

\begin{eqnarray}
\frac{\mu}{t} = \left\{
  \begin{array}{ll}
    2 \cos\biggr(\frac{\pi n}{f(u)}\biggr), & \hbox{$0 \leq n <1$} \\
     u + 2 \cos\biggr(\frac{\pi (2-n)}{f(u)}\biggr), & \hbox{$1 < n \leq 2$}.
  \end{array}
\right. \end{eqnarray}

\noindent The average particle densities are found by inverting this equation,
\begin{widetext}
\begin{eqnarray}
n = \left\{
  \begin{array}{ll}
    0, & \hbox{$\mu/t < -2$;} \\
    \frac{f(u)}{\pi}\arccos[-\mu/2t], & \hbox{$-2 \leq \mu/t \leq -2 \cos[\pi/f(u)]$} \\
    1, & \hbox{$-2 \cos[\pi/f(u)] < \mu/t < u + 2 \cos[\pi/f(u)]$} \\
    2 - \frac{f(u)}{\pi} \arccos[(\mu/t-u)/2], & \hbox{$u + 2 \cos[\pi/f(u)] \leq \mu/t \leq u + 2$} \\
    2, & \hbox{$\mu/t > u + 2$.}
  \end{array}
\right.\end{eqnarray}
\end{widetext}

\noindent For a representative value of the on-site interaction $U/t = 5$, the zero temperature particle density as a function of the chemical potential is shown in FIG.~\ref{f1}. As we stated before, the chemical potential-axis can be considered as a spatial coordinate where the minimum chemical potential is at the edge of the trap. As can be seen from the figure, five phases- two metallic phases, two insulating phases, and a vacuum phase- can coexist inside the trap if the total number of particles in the trap is large. While density pins at 1 in the Mott insulating phase, it pins at 0 and 2 in the vacuum and band insulating phases, respectively.

 \begin{figure}
\includegraphics[width=\columnwidth]{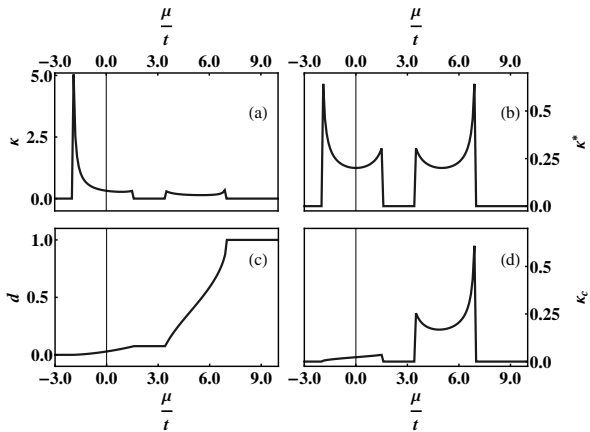}
\caption{Zero-temperature compressibilities and double occupation as a function of chemical potential $\mu$. The interaction is fixed at $U/t = 5$.} \label{f2}
\end{figure}

\begin{figure}
\includegraphics[width=\columnwidth]{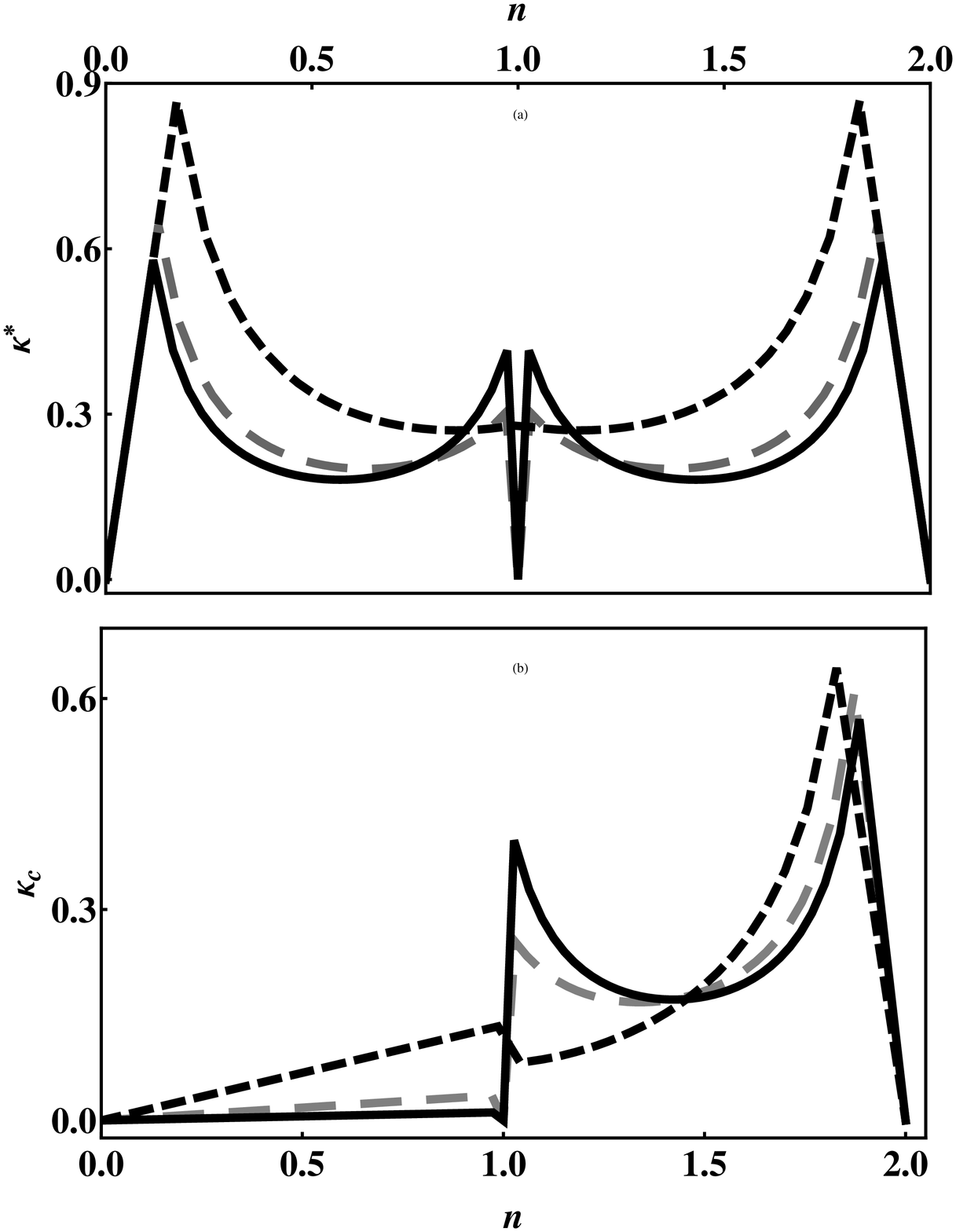}
\caption{Zero-temperature compressibilities as a function of the particle density for three representative values of interactions. long dashed (grey) line: $U/t = 1$, short dashed line: $U/t = 5$, solid line: $U/t =10$.} \label{f3}
\end{figure}

We calculate two compressibilities $\kappa^* = \partial n/\partial \mu$ and $\kappa = \kappa^*/n$. Panels (a) and (b) of FIG.~\ref{f2} show these compressibilities as a function of chemical potential. The sharp peak of $\kappa$ at the edge of the trap is because of the low density. While this sharp peak is absent in $\kappa^*$, it shows two compressible
regions corresponding to the lower metallic band and the upper metallic band. Further, we calculate the double occupation
$d = \sum_i \langle n_{i\uparrow}n_{i\downarrow} \rangle = \partial e/\partial U$ and the \emph{double-occupancy} compressibility $\kappa_c = \partial d/\partial \mu$ which are shown in panels (c) and (d) of FIG.~\ref{f2}. Notice that the panel (b) must be compared with the panel (d) as both represent the unnormalized compressibilities. Because of the sharp peak in the normalized compressibility $\kappa$, we do not attempt to compare the normalized \emph{double-occupancy} compressibility. Although we have fixed the on-site interaction $u$ to be 5, the qualitative behavior would be the same for all larger values of $u$. The three incompressible regions in panels (a), (b), and (d) corresponding to a
pinning of the density at $n$ =0, 1, and 2. Notice that the \emph{double-occupancy} compressibility $\kappa_c$ measures only the core region [panel (d)] so that it
is \emph{not} sensitive to the edge of the cloud. Using a high temperature expansion Scarola \emph{et al}.~\cite{t5} have calculated the compressibility $\kappa$ and the \emph{double-occupancy} compressibility $\kappa_c$ for a three dimensional optical lattice. In the next section, we show that our high temperature compressibilities in 1D also show very similar characteristic behavior as those of the calculation in Ref.~\cite{t5}. In the insulating phases the compressibilities are zero so they are discontinuous at densities $n =0, 1$, and $2$. These characteristic properties are directly related to the formation of gaps in the insulating phases. As we will see later, these discontinuities smear out as the temperature increases. As shown in panel (d) of FIG.~\ref{f2}, the peak corresponding to the lower band disappears in $\kappa_c$ so that the \emph{double-occupancy} compressibility reveals the double occupancy and the Mott insulator transition in the center of the trap. The measurements of both double occupation and a quantity related to compressibilities, $\partial d/\partial N = \kappa_c/\kappa^*$ for a three dimensional lattice were made in Ref~.\cite{fermi1}. In FIG.~\ref{f3}, we compare $\kappa^*$ and $\kappa_c$ for different values of on-site interactions $u$. When the chemical potential is large, the upper band is occupied so that the system has doubly occupied sites. For this case, $\kappa_c$ is identical to $\kappa^*$ at chemical potentials corresponding to the upper band, but it is almost zero at the lower band.

\medskip
\section{IV. Finite temperature results}

At finite temperatures, we numerically solve TBA equations to find compressibilities and entropy. Following Takahashi~\cite{takahashi}, the thermodynamic potential per site is given by

\begin{widetext}

\begin{eqnarray}
\Omega = e_0-\mu -k_BT \biggr\{\int^\pi_{-\pi} \rho_0(k) \ln[1+\xi(k)] dk + \int^\pi_{-\pi}\sigma_0(\Lambda) \ln[1+\eta_1(\Lambda)] d\Lambda \biggr\}.
\end{eqnarray}
\end{widetext}

\noindent The energy per site $e_0 = 2tI$, and two distribution functions of $k$'s and $\Lambda$'s are given by~\cite{fn},

\begin{eqnarray}
\rho_0(k) &=& \frac{1}{2\pi} + \cos k\int^{\infty}_{-\infty}a_1(\Lambda-\sin k)\sigma_0(\Lambda) d\Lambda \\ \nonumber
\sigma_0(\Lambda) &=& \frac{1}{2\pi}\int^{\infty}_{-\infty}s(\Lambda-\sin k)dk.
\end{eqnarray}

\noindent Here $a_1(x) = 4 u/[\pi (u^2 + 16 x^2)]$ and $s(x) = \csc (2\pi x/u)/u$. The particle-hole ratios of $k$ excitations and $\Lambda$ excitations, $\xi(k)$ and $\eta_1(k)$ are obtained by an infinite set of nonlinear integral equations:

\begin{widetext}
\begin{eqnarray}
\ln\xi(k) &=& \frac{\kappa_0(k)}{T} + \int_{-\infty}^\infty d\Lambda s(\Lambda-\sin k) \ln\biggr(\frac{1+\eta_1^\prime(\Lambda)}{1+\eta_1(\Lambda)}\biggr) \\
\ln\eta_1(\Lambda)&=& s^*\ln[1+\eta_2(\Lambda)]-\int^\pi_{-\pi} s(\Lambda-\sin k) \ln[1+\xi^-1(k)]\cos k dk \\ \nonumber
\ln\eta_1^\prime(\Lambda)&=& s^*\ln[1+\eta_2^\prime(\Lambda)]-\int^\pi_{-\pi} s(\Lambda-\sin k)\ln[1+\xi(k)]\cos k dk \\ \nonumber
\end{eqnarray}

\noindent and for $j \geq 2$,

\begin{eqnarray}
\ln\eta_j(\Lambda)&=& s^*\ln\{[1+\eta_{j-1}(\Lambda)][1+\eta_{j+1}(\Lambda)]\} \\
\ln\eta_j^\prime(\Lambda)&=& s^*\ln\{[1+\eta_{j-1}^\prime(\Lambda)][1+\eta_{j+1}^\prime(\Lambda)]\}. \\ \nonumber
\end{eqnarray}
\end{widetext}

\noindent Here $s^*f(\Lambda)\equiv\int^\infty_{-\infty}s(\Lambda-\Lambda^\prime)f(\Lambda^\prime)d\Lambda^\prime$ and $\kappa_0(k) \equiv -2t\cos k-4t\int^\infty_{-\infty}d\Lambda s(\Lambda-\sin k) \times Re\sqrt{1-(\Lambda-u i/4)^2}$. In order to calculate the thermodynamic potential numerically, one has to cut off the set of infinite equations at a finite number $j$. We achieve this by following the numerical procedure proposed by Takahashi \emph{et al}.~\cite{takahashiN}. The infinite set of equations is truncated by replacing $s(\Lambda)$ by $\delta(\Lambda)/2$ at $j > n_c$. Then the integral equations are converted into a set of matrix equations in which $2n_c +1$ unknown functions are represented in terms of discrete points of $k$ and $\Lambda$. These non linear matrix equations are then solved iteratively using Kepler's method. The details of the numerical procedure can be found in Ref.~\cite{takahashiN}. From the numerical solutions of the non-linear integral equations, we calculate thermodynamic potential $\Omega$, particle density $n = -\partial \Omega/\partial \mu$, compressibility $\kappa^* = \partial n/\partial \mu$, entropy $S= -\partial \Omega/\partial T$, and \emph{double-occupancy} compressibility $\kappa_c = \partial d/\partial \mu$, where $d = \sum_i \langle n_{i\uparrow}n_{i\downarrow} \rangle = \partial \Omega/\partial U$ is the double occupation.

\begin{figure}
\includegraphics[width=\columnwidth]{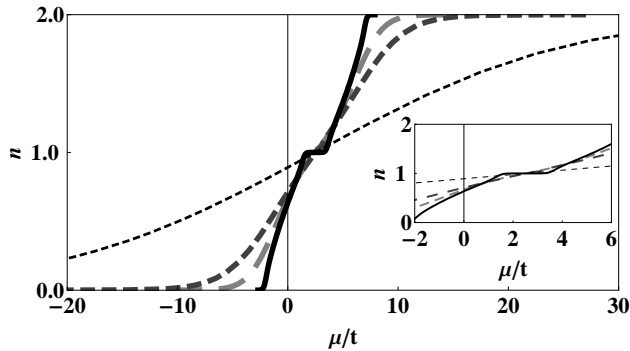}
\caption{Particle Density as a function of the chemical potential at $U/t =5$. The four lines ranging from solid to short dashed represent different temperatures $(k_BT)/t$ = 0.1, 1, 2, and 10 respectively.} \label{f5}
\end{figure}

\begin{figure}
\includegraphics[width=\columnwidth]{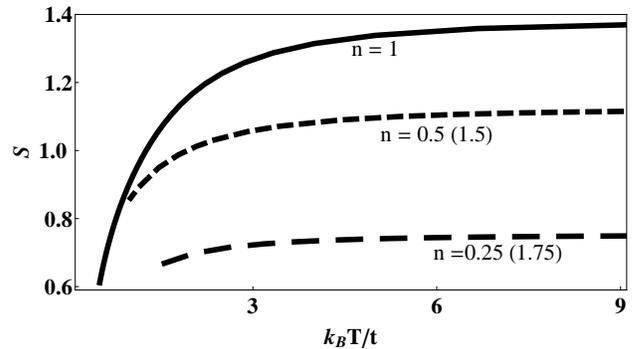}
\caption{The temperature dependence on entropy and particle density at $U/t=5$. Because of particle-hole symmetry, the entropy at $n = 0.5 (0.25)$ is equal to the entropy at $n = 1.5 (1.75)$.} \label{f6}
\end{figure}

First, we present particle density as a function of chemical potential at different temperatures. As shown in FIG.~\ref{f5}, the plateau region at half filling smears out as the temperature increases. However at very low temperatures, the plateau region still survives showing a Mott insulating gap at $n =1$. The calculated entropy at three different filling factors is shown in FIG.~\ref{f6}. In general, the entropy monotonically increases with temperature, however, the increase is very strong at lower temperatures. In Ref.~\cite{neel3D}, the entropy at half filling for 3D optical lattice systems is calculated focusing on the anti-ferromagnetic (AFM) transition into Neel state. Even for a 2D \emph{finite} lattice, a low temperature crossover scale is identified as a growth of AFM order~\cite{neel2D}. As the finite temperature Neel transition is absent in 1D systems and the system under consideration is in thermodynamic limit, we do not find a crossover scale in our 1D calculation.

\begin{figure}
\includegraphics[width=\columnwidth]{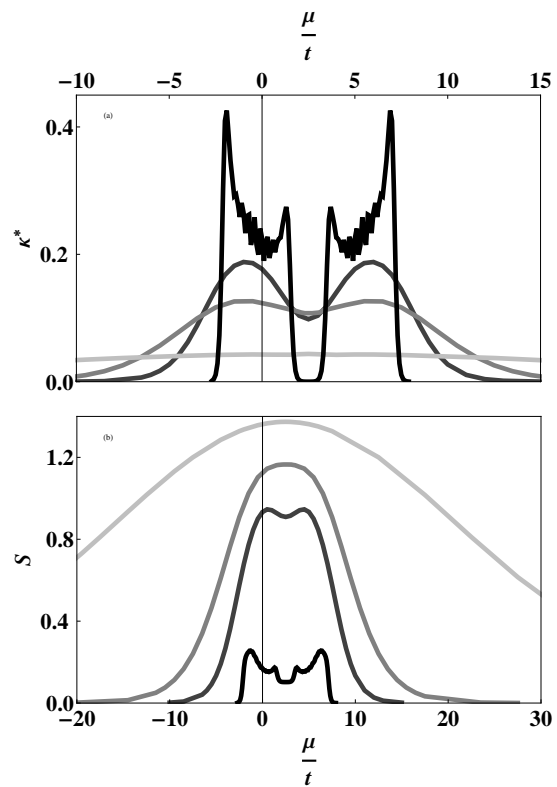}
\caption{Compressibility and entropy at $U/t =5$. The four lines ranging from dark to light colors represent different temperatures $(k_BT)/t$ = 0.1, 1, 2, and 10 respectively.} \label{f7}
\end{figure}

The compressibility $\kappa^*$ and the entropy $S$ at fixed temperatures are shown in FIG.~\ref{f7}. As can be seen, the double peak structure in $\kappa^*$ and $S$ disappear at higher temperatures. The disappearance of these sharp features may be used as a probe of temperature in experiments. The enhancements of compressibility contrast to the suppressions of entropy at lower temperatures are attributed to the metal-insulator transition at the center of the trap. At sufficiently high temperatures, the effect of interaction is weak. As the temperature decreases, the interaction leads to the opening of an Mott gap in the excitation spectrum at half filling. The low temperature results of FIG.~\ref{f7} clearly indicate this characteristic feature. The wiggles in $\kappa^*$ at $k_BT = 0.1t$ are due to the numerical noises in our low temperature calculation; however, the qualitative feature resembles the zero temperature results. Notice that the entropy is largest in the vicinity of half filling at higher temperatures, it is larger in metallic phases at lower temperatures. This interesting feature opens the possibility of using entropy measurements to detect th metal-insulator transition at the center of the trap, as we see in the next section.

\begin{figure}
\includegraphics[width=\columnwidth]{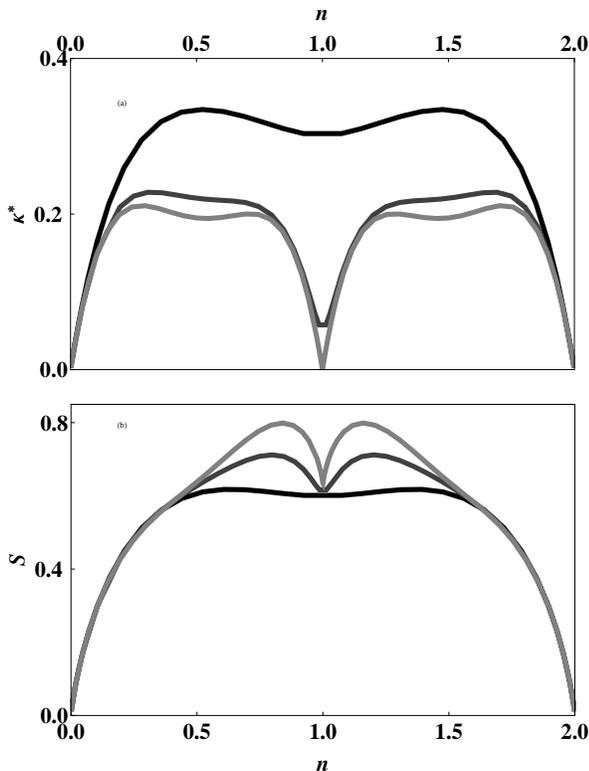}
\caption{Compressibility and entropy as a function of particle density at $k_BT$ = 0.5 t. The three lines ranging from dark to light colors represent different interactions $U/t$ = 1, 5, and 10 respectively.} \label{f8}
\end{figure}

In FIG.~\ref{f8}, we present the compressibility $\kappa^*$ and entropy $S$ at a fixed temperature for different interactions involving different $u$. As expected, the compressibility decreases with increasing interaction and it is at a minimum in insulating phases. The interaction significantly increases the entropy at the vicinity of the Mott insulating phase (in the regions $0.5 \leq n < 1$ and $1 < n \leq 1.5$). However, deep in the Mott insulating phase, the interaction has a very weak effect on the entropy. This is because the interaction increases the excitation gap and the density of low-energy states. For larger interactions, the excitation gap scales linearly with the interaction, but it is exponentially small for smaller interactions.

\begin{figure}
\includegraphics[width=\columnwidth]{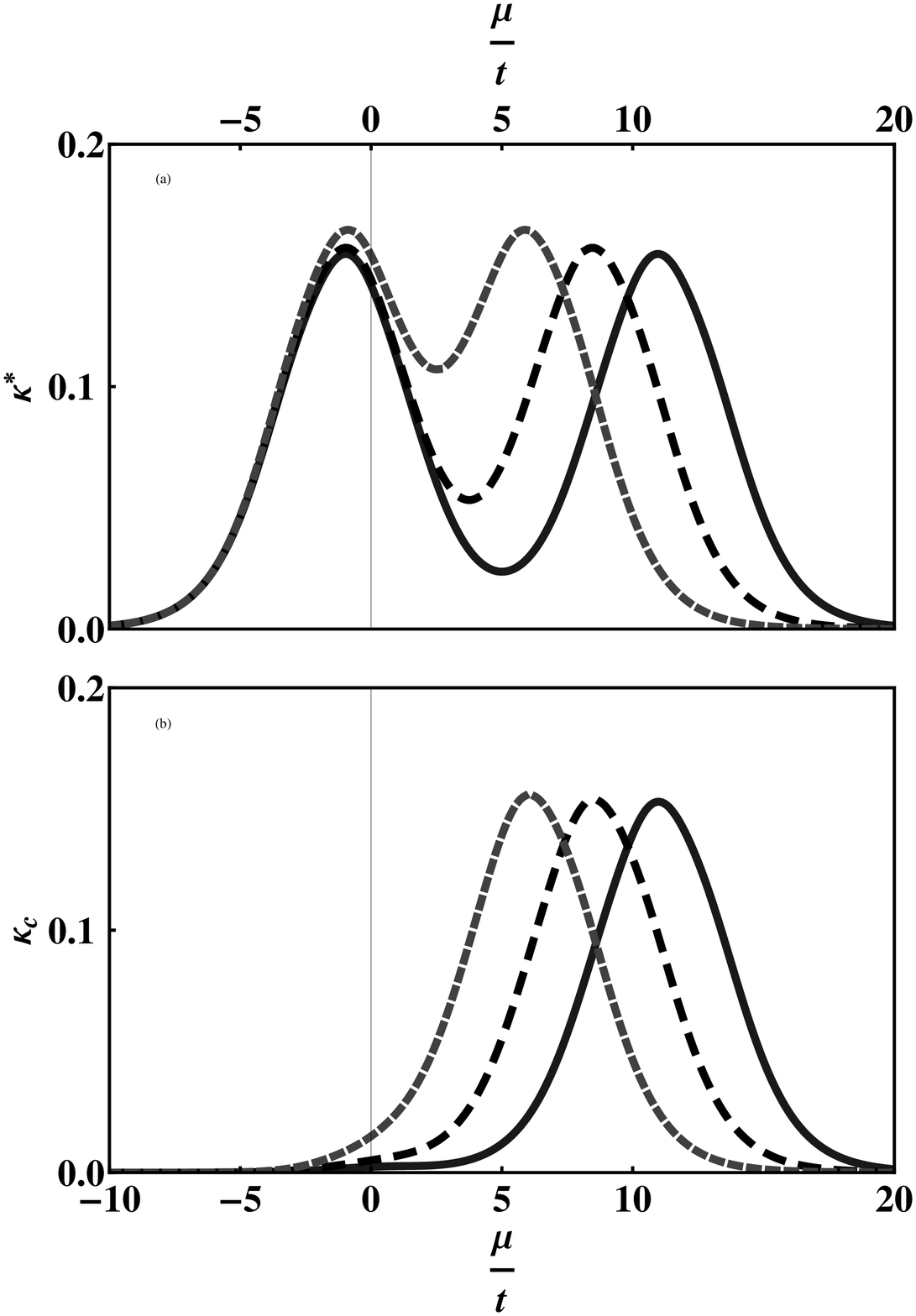}
\caption{Compressibility and double-occupancy compressibility as a function of the chemical potential at $k_BT$ = 1.33 t. The three lines, solid, short dashed (black), and long dashed (gray) represent different interactions $U/t$ = 10, 7.5, and 5 respectively.} \label{f9}
\end{figure}

The compressibility $\kappa^*$ and \emph{double-occupancy} compressibility $\kappa_c$ are shown in FIG.~\ref{f9}. As we have seen in zero temperature results, both $\kappa$ and $\kappa_c$ are almost identical at chemical potentials corresponding to the upper metallic band, but $\kappa_c$ is zero at lower chemical potentials corresponding to the lower metallic band. A high temperature expansion method in three dimensions carried out by Scarola \emph{et al}.~\cite{t5} also shows very similar qualitative behavior as our 1D exact results. As we discussed before, the occupation of the upper band is directly observable through the measurements of \emph{double-occupancy} compressibility. As can be seen from the figure, the upper band moves toward larger chemical potentials as the interaction increases. Again, this is because of the increase of the excitation gap with interaction.

\section{V. Connections to Experiments}

This section is devoted to a discussion of the connection of our calculations to the experimental data. The density variations which can be measured using \emph{in situ} images, can be used to deduce the bulk thermodynamic properties within the local density approximation (LDA). In order to study 1D-lattice fermions in laboratories, an array of 1D-lattice geometries is formed with a 3D optical lattice. Apart from the underlying harmonic potential, the lattice potential has the form $V = V_{0\perp}[\cos^2(kx)+\cos^2(ky)]+V_{0z}\cos^2(kz)$, with $k=2\pi/\lambda$, where $\lambda$ is the optical-trap laser wavelength. For the system to be in a 1D regime, the transverse laser intensity $V_{o\perp}$ must be much larger than the axial laser intensity $V_{0z}$ so that the transverse tunneling rate is negligible compared to the axial tunneling rate. Further, the transverse confinement $\hbar\omega_\perp >> k_Bt$, $\epsilon_F$ so that only the lowest transverse mode in each lattice tube is populated.

The LDA assumes that the atomic system is locally homogeneous with local chemical potentials given by $\mu_{i} = \mu_0-V_i$. Here $\mu_0$ is the chemical potential at the center of the trap. Then combining this chemical potential with the experimentally measured column density profiles [$n(y, z)$]of the 1D-lattice bundles, the thermodynamic quantities can be extracted. The first step is the construction of the linear axial density $n(z) = \int n(y, z) dy$. For non-lattice 1D systems, these linear density profiles have been already measured at Rice University~\cite{rice}. Notice that the compressibilities $\kappa$ and $\kappa^*$ are related to the density $n(z)$ and $\partial n/\partial \mu$. Since $\partial n/\partial \mu \propto \partial n/\partial z$, these compressibilities can be extracted from the column densities. The extraction of the \emph{double-occupancy} compressibility $\kappa_c = \partial d/\partial \mu$ involves the calculation of $\partial d/\partial N$. As presented in Ref.~\cite{fermi1}, this quantity can be extracted from the double occupancy. Then using the relations $\partial d/\partial N = (\partial d/\partial \mu) (\partial \mu/\partial N)$ and $\partial N/\partial \mu = (\partial N/\partial z) (\partial \mu/\partial z)^{-1}$, the \emph{double-occupancy} compressibility can be extracted from experimental data.

The entropy $S = (\partial P/\partial T)_{\mu}$ needs two different configurations of pressure $P$ at two different temperatures $T$, but at the same chemical potential $\mu$. Following the proposal by Ho  and Zhou~\cite{ho}, the local pressure $P(z) = m\omega^2_\perp n(z)/(2\pi)$ can be extracted as for a 3D gas~\cite{ens}. Two different column density profiles can be measured at two different axial trapping frequencies $\omega_{z1}$ and $\omega_{z2}$, where $\omega_{z2} \simeq \omega_{z1}$. After taking an \emph{in situ} image at $\omega_{z1}$, and then adiabatically changing $\omega_{z1} \rightarrow \omega_{z2}$, a second image can be taken. During this adiabatic change in the trapping frequency, both temperature and chemical potential will slightly change. Then calculating the pressure at two different positions such that $\mu_{z1} = \mu_{z2}$, the entropy $S(z) = [P(z2)-P(z1)]/(T_2-T_1)$ can be extracted, where $T_1$ and $T_2$ are the temperatures of the system at the axial trapping frequencies $\omega_{z1}$ and $\omega_{z2}$ respectively~\cite{ho, qi}.

So far the present discussion has been about how to extract the homogeneous thermodynamic quantities from the experimental measurements for tapped atoms. Let us now turn to the calculation of the entropy and the compressibility in harmonic traps so that our calculation can be compared with the experimentally determined total entropy and compressibility of trapped Fermi gases. Consider the total number of atoms $N = \int dz n(z)$. Within the local density approximation, the integration over $z$ can be converted to an integration over the local chemical potential $d\mu = -m\omega z dz$ and we find

\begin{eqnarray}
N = \sqrt{\frac{2t}{m\omega^2}}\int^{\tilde{\mu}_0}_{-\infty}\frac{n(\tilde{\mu})}{\sqrt{\tilde{\mu}-\tilde{\mu}_0}}d\tilde{\mu}.
\end{eqnarray}

\noindent Here $\omega$ is the one-dimensional trapping frequency and $\tilde{\mu} =\mu/t$. The central chemical potential $\tilde{\mu}_0 = \mu_0/t$ is determined by the total number of atoms in the trap. Using the same change of variable, we find the total entropy and the total compressibility,

\begin{eqnarray}
S_T = \sqrt{\frac{2t}{m\omega^2}}\int^{\tilde{\mu}_0}_{-\infty}\frac{S(\tilde{\mu})}{\sqrt{\tilde{\mu}-\tilde{\mu}_0}}d\tilde{\mu} \\ \nonumber
\kappa^*_T = \sqrt{\frac{2t}{m\omega^2}}\int^{\tilde{\mu}_0}_{-\infty}\frac{\kappa^*(\tilde{\mu})}{\sqrt{\tilde{\mu}-\tilde{\mu}_0}}d\tilde{\mu}.
\end{eqnarray}

\begin{figure}
\includegraphics[width=\columnwidth]{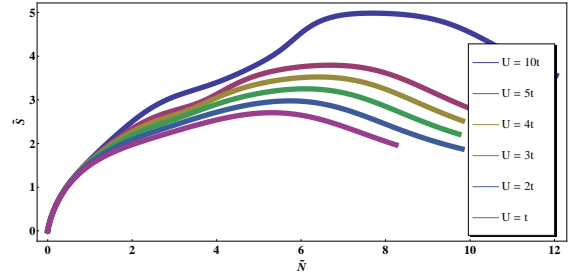}
\caption{[Color online]Total entropy as a function of the total number of atoms in a harmonic trap.} \label{st}
\end{figure}

\begin{figure}
\includegraphics[width=\columnwidth]{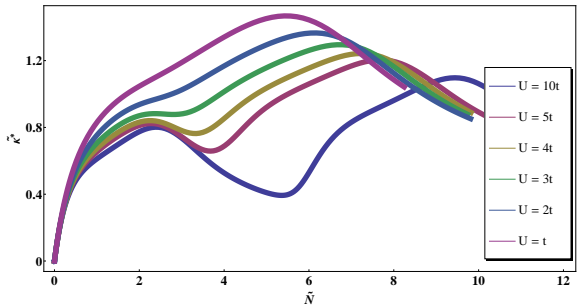}
\caption{[Color online]Total compressibility as a function of the total number of atoms in a harmonic trap.} \label{kt}
\end{figure}

\begin{figure}
\includegraphics[width=\columnwidth]{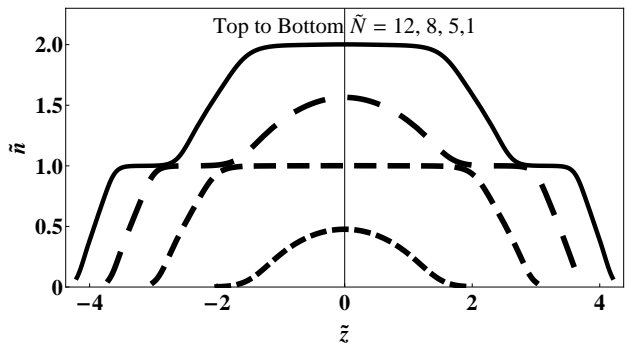}
\caption{Density profiles for different atom numbers in the 1D lattice for $U/t = 10$.} \label{df10}
\end{figure}

\noindent Figures~\ref{st} and~\ref{kt} respectively show the entropy $\tilde{S} = S_T \sqrt{m\omega^2/2t}$ and $\tilde{\kappa^*} = \kappa^*_T \sqrt{m\omega^2/2t}$ as functions of the total number of particles $\tilde{N} = N \sqrt{m\omega^2/2t}$ for different interactions $U/t$ at a fixed temperature $k_BT = 0.5$. Both the entropy and compressibility clearly show the appearance of insulating and metallic phases in the trap as we increase the number of atoms or the interaction strength. As an example, we plot the density profiles as a function of $\tilde{z} = m\omega^2 z/2t$ in FIG.~\ref{df10} for the case of $U = 10 t$. These density profiles correspond to different numbers of atoms $N$. Notice that each maximum of the compressibility indicates the existence of two metallic phases (upper band and lower band) while each minimum indicates the existence of two insulating phases (Mott insulator and band insulator) inside the cloud. For example, when $\tilde{N} = 1$, the compressibility does not show any minima so that the entire cloud is metallic. For the case of $\tilde{N} = 8$, the compressibility shows two maxima corresponding to the two metallic phases separated by the Mott insulator phase. Though it is very weak, the same behavior can be seen in the entropy profile too.

\noindent The total double occupation $D_T = \int dz d(z)$ has the form

\begin{eqnarray}
D_T = \sqrt{\frac{2t}{m\omega^2}}\int^{\tilde{\mu}_0}_{-\infty}\frac{d(\tilde{\mu})}{\sqrt{\tilde{\mu}-\tilde{\mu}_0}}d\tilde{\mu}.
\end{eqnarray}

\begin{figure}
\includegraphics[width=\columnwidth]{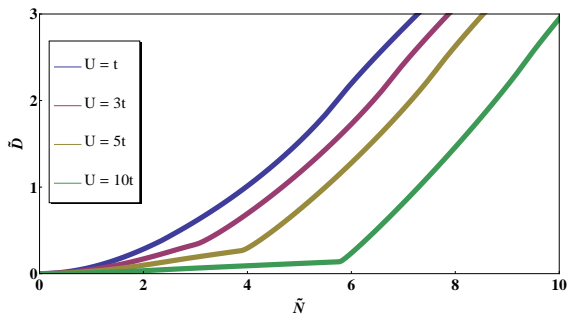}
\caption{[Color online] Total number of double occupation sites in a harmonic trap.} \label{dt}
\end{figure}

\noindent The zero temperature total double occupation $\tilde{D} = D_T\sqrt{m\omega^2/2t}$ as a function of $\tilde{N}$ for different interactions is shown in FIG.~\ref{dt}. As one expects, the double occupation increases with the number of atoms but it is suppressed with increasing interaction.

\section{VI. Discussion and conclusions}

We have calculated several thermodynamic quantities for 1D fermions in optical lattices. Modeling the fermions in the lattice by the 1D Hubbard model, we have calculated the particle density, compressibility, and entropy by means of the Bethe ansatz method. We find that these quantities have characteristic features that depend on the chemical potential, temperature, and interaction. As one expects, the sharp features on these quantities at low temperatures become smooth at higher temperatures and the visible signs of the Mott plateaus disappear. Our compressibility calculation shows that one can use compressibility measurements to detect stable compressible phases. In particular, we find that interaction effects, double occupation effects and temperature effects can be probed through measurements of the \emph{double-occupancy} compressibility without the effect of trap edges. At relatively higher temperatures, while the compressibility $\kappa^* = \partial n/\partial \mu$ is finite in the entire range of the atomic cloud, the \emph{double-occupancy} compressibility $\kappa_c = \partial d/\partial \mu$ is finite only in the upper band. Therefore, the \emph{double-occupancy} compressibility can be used to detect the occupation of the upper band at the center of the trap. Further, we find that entropy also shows characteristic features; at higher temperatures entropy is larger at half filling, at low temperatures it is larger at the lower and the upper bands. Therefore, entropy measurements would allow one to monitor the metal-insulator transition at the center of the trap. We use the local density approximation to take into account the effect of the underlying harmonic trapping potential and find that measurements of entropy and compressibility of the entire cloud reveal the detailed structure (namely the existence of insulating and/or metallic phases) of the trapped atoms in the lattice.

In current cold-atom experiments, entropy can be monitored more easily than temperature. The lowest achieved entropy per particle is $S \simeq \ln 2$. Our calculations (FIG.~\ref{f8}) show that entropy at half filing is approximately equal to $\ln 2$. Therefore, the current lowest temperature experiments should allow one to observe the insulating gap and the metal-insulator transition. One dimensional lattice fermions with an attractive interaction are analogous to the half-filling repulsive Hubbard model with a finite magnetization. Therefore, the present work can be generalized to an attractive Hubbard system by introducing a population imbalance.

\section{ACKNOWLEDGMENTS}

We are very grateful to Patrick O'Brien for discussions and critical comments on the manuscript.


\begin{references}
\bibitem{fn} A good discussion on TBA equations can be found, for example, in chapter 14 of reference~\cite{tak1}.
\end{references}
\end{document}